\documentclass[sn-mathphys]{sn-jnl}

\usepackage{graphicx}%
\usepackage{multirow}%
\usepackage{multicol}%
\usepackage{amsmath,amssymb,amsfonts}%
\usepackage{amsthm}%
\usepackage{mathrsfs}%
\usepackage[title]{appendix}%
\usepackage{xcolor}%
\usepackage{textcomp}%
\usepackage{manyfoot}%
\usepackage{booktabs}%
\usepackage{algorithm}%
\usepackage{algorithmicx}%
\usepackage{algpseudocode}%
\usepackage{listings}%
\usepackage{subfigure}
\usepackage{float} 
\usepackage{caption}
\usepackage{natbib}

\DeclareMathOperator*{\argmin}{arg\,min}

\graphicspath{{Figures/}} 

\theoremstyle{thmstyleone}%
\theoremstyle{thmstyletwo}%

\theoremstyle{thmstylethree}%

\begin{document}

\title[Plug-and-Play regularized 3D seismic inversion with 2D pre-trained denoisers]{Plug-and-Play regularized 3D seismic inversion with 2D pre-trained denoisers}


\author*[1]{\fnm{Nick} \sur{Luiken}}\email{nicolaas.luiken@kaust.edu.sa}

\author[1]{\fnm{Juan} \sur{Romero}}\email{juan.romeromurcia@kaust.edu.sa}

\author[1]{\fnm{Miguel} \sur{Corrales}}\email{miguel.corrales@kaust.edu.sa}

\author[1]{\fnm{Matteo} \sur{Ravasi}}\email{matteo.ravasi@kaust.edu.sa}

\affil*[1]{\orgdiv{Earth Science and Engineering, Physical Sciences and Engineering (PSE)}, \orgname{King Abdullah University of Science and Technology (KAUST)}, \orgaddress{\city{Thuwal}, \postcode{23955-6900}, \country{Kingdom of Saudi Arabia}}}



\abstract{Post-stack seismic inversion is a widely used technique to retrieve high-resolution acoustic impedance models from migrated seismic data. Its modelling operator assumes that a migrated seismic data can be generated from the convolution of a source wavelet and the time derivative of the acoustic impedance model. Given the band-limited nature of the seismic wavelet, the convolutional model acts as a filtering operator on the acoustic impedance model, thereby making the problem of retrieving acoustic impedances from seismic data ambiguous. In order to compensate for missing frequencies, post-stack seismic inversion is often regularized, meaning that prior information about the structure of the subsurface is included in the inversion process. Recently, the Plug-and-Play methodology has gained wide interest in the inverse problem community as a new form of implicit regularization, often outperforming state-of-the-art regularization. Plug-and-Play can be applied to any proximal algorithm by simply replacing the proximal operator of the regularizer with any denoiser of choice. We propose to use Plug-and-Play regularization with a 2D pre-trained, deep denoiser for 2D post-stack seismic inversion. Additionally, we show that a generalization of Plug-and-Play, called Multi-Agent Consensus Equilibrium, can be adopted to solve 3D post-stack inversion whilst leveraging the same 2D pre-trained denoiser used in the 2D case. More precisely, Multi-Agent Consensus Equilibrium combines the results of applying such 2D denoiser in the inline, crossline, and time directions in an optimal manner. We verify the proposed methods on a portion of the SEAM Phase 1 velocity model and the Sleipner field dataset.}




\maketitle

\section{Introduction}\label{sec1}

Characterizing the Earth's subsurface is of paramount importance for many energy applications ranging from oil and gas exploration, hydrocarbon production, carbon capture utilization and storage (CCUS), and assessment of construction sites for wind farms, just to name a few. Various remote sensing techniques are usually employed to estimate different properties of the subsurface, with reflection seismic being one of the most commonly employed methods given its sensitivity to rock and fluid properties. A direct relation can be found between the subsurface properties and the recorded seismic data via the physics of the elastic wave propagation. However, due to complex propagation phenomena (e.g., scattering effects), such a relation is highly non-linear; the difficulty in estimating the subsurface properties is further exacerbated by the use of limited acquisition setups, where the subsurface is probed from only one side with limited aperture arrays of sources and receivers. To mitigate such non-linearity, the relation between seismic data and the subsurface is often described by means of simplified models that assume, for example, a layered subsurface or acoustic wave propagation. The convolutional model \citep{Goupillaud1961} represents one example of such simplified models, as it provides a direct, linear relationship between migrated seismic data and acoustic properties of the subsurface; the migrated seismic data can, in fact, be represented by the convolution of a given wavelet with the acoustic impedance model. Retrieving such an acoustic impedance model based on the knowledge of the wavelet and the migrated seismic data entails solving an inverse problem, commonly dubbed in the literature under the name of post-stack seismic inversion \citep{Oldenburg1983, Russell1991}. However, given the band-limited nature of the seismic wavelet, the relation between the acoustic impedance and the seismic data is non-unique, meaning that multiple acoustic impedance models could equally describe the same seismic data, rendering the inverse problem ill-posed.

Regularization is a widely used method to deal with the non-uniqness of ill-posed inverse problems \citep{Engl1996}. It refers to the process of adding prior information when solving an inverse problem, typically encoded through an additional term in the objective function alongside the data misfit term. This additional term is tasked to compensate for the missing information in the data, i.e. information that is lost in the generation of the data because of the nullspace of the underlying physical operator; in the case of post-stack seismic inversion, this is represented by the missing low and high frequencies, which are outside of the bandwidth of the seismic wavelet. One of the earliest forms of regularization, known as Tikhonov regularization \citep{Tikhonov1977}, consists of adding an $\ell_2$-penalty term to the objective function, thereby damping high amplitude values in the solution that can arise when inverting the operator of an ill-posed inverse problem. A practical benefit of Tikhonov regularization is that it renders an objective function is that is easy to minimize, since the $\ell_2$ norm is convex and differentiable. An alternative to $\ell_2$-norm regularization was later proposed in the form of $\ell_1$-norm regularization \citep{Mallat1993, Chen1994, Daubechies2004}, also known as sparse regularization, which enforces the solution to be sparse (i.e., few non-zero coefficients). Although early applications of this concept exist in the literature \citep{Santosa1986}, a factor that limited the widespread adoption of the $\ell_1$-norm regularization is the fact that the resulting objective function becomes non-differentiable. As such, $\ell_1$ problems had to be solved using linear programming methods, which are notoriously slow to converge \citep{Donoho1989}. More recently, new algorithms have emerged to efficiently solve $\ell_1$ problems: these advances have been mostly led by the application of sparse inversion in compressed sensing problems \citep{Candes2006a}, where a signal is reconstructed from measurements sampled below the Nyquist rate.Reconstruction is guaranteed under certain conditions of the sampling matrix and provided an $\ell_1$-norm penalty is applied to the signal of interest or to a suitable transformation of such signal \citep{Candes2006b}. For example, throughout the years, sparsity has been applied in conjunction with a variety of linear transformations, such as the one- or multi-dimensional Fourier transform \citep{Candes2006b}, the Wavelet transform \citep{Daubechies1990, Daubechies2004}, or the Curvelet transform \citep{Starck2002, Herrmann2008}. Another instance of sparse regularization is Total-Variation (TV) regularization \citep{Rudin1992}, which applies a sparse penalty on the gradient of the model parameters, thereby enforcing piece-wise constant solutions. This is especially suitable for subsurface models \citep{Gholami, Kolbjornsen, Ravasi2022}, which is motivated by the fact that the Earth presents a layered structure with a preferential direction of smoothness (i.e., along key geological horizons) and an orthogonal direction of blockiness (i.e., across key geological horizons).

Proximal algorithms \citep{Parikh2014, Eckstein1992, Rockafellar1976} have recently proven to be effective for solving inverse problems involving $\ell_1$ constraints. In essence, proximal algorithms introduce one or more auxiliary variables that decouple different terms in the objective function, allowing the original problem to be solved in an alternating fashion for one variable at the time. Commonly, the variable(s) associated with smooth terms (e.g., the data misfit term) is updated by one or a few gradient steps, whilst the other variable(s) is updated using the so-called proximal operator; in the case of the $\ell_1$ norm (as well as many other popular norms used in inverse problem theory), the proximal operator has a closed-form solution, making proximal algorithms very efficient and flexible. Recently, a new class of method called Plug-and-Play (PnP) algorithms \citep{Venkatakrishnan2013} have gained popularity in the inverse problem community; as we will see later in the text, such algorithms are specialized versions of proximal algorithms, obtained from the realization that the proximal operator of any function can be interpreted as a Maximum-A-Posterior (MAP) denoising problem. This realization allowed the original authors to substitute the proximal operator of any regularization term with any denoiser of choice (which may or may not have an associated objective function to minimize). Initially, the denoisers of choice were statistical denoisers such as BM3D \citep{Dabov2007} and non-local means \citep{Buades2005}; however, with the rise of deep learning, neural-network based denoisers became state-of-the-art \citep{Zhang2017b}. One of the issues with Plug-and-Play algorithms is that by replacing the proximal operator with a denoiser, the coupling with an underlying objective function is lost, and due to the fact that the problem is no longer convex, all of the proven convergence guarantees of proximal algorithms cannot be used. A variety of works have studied conditions under which PnP converges \citep{Ryu2019}; however, in order to do so, such studies place restrictive assumptions on the network architecture and the physical modelling operator, which cannot always be satisfied in real life applications. An appealing alternative to PnP methods is represented by the Multi-Agent Consensus Equilibrium (MACE -- \cite{Buzzard2018}) algorithm. Rather than minimizing an objective function, MACE simply consists of a number of agents, which could be proximal operators, denoisers, or any other function; the algorithm requires such agents to work together until consensus equilibrium is reached, meaning that each agent produces the same output and that such output does not change if further iterations are carried out. This framework allows for a more natural combination of physical operators and denoisers since no underlying objective function is needed. 

\paragraph{Our contribution}
Our contribution is two-fold:
\begin{enumerate}
    \item we frame the post-stack seismic inverse problem as a PnP regularized inversion that leverages a state-of-the-art 2D pre-trained deep denoiser, and show that it outperforms TV regularization on a section of the 3D SEAM Phase 1 velocity model;
    \item we propose the use of consensus optimization for 3D post-stack seismic inversion by applying a single 2D pre-trained deep denoiser over different directions of the model (i.e., inline, crossline, and time). We show that by combining the physics of the post-stack modelling operator and the power of 2D deep denoisers within the consensus equilibrium framework leads to high-quality results that once again outperform state-of-the-art TV-regularization. 
\end{enumerate}

\section{Theory}
\subsection{Post-stack seismic inversion}
Reflection seismology \citep{Sheriff} aims at characterizing the subsurface by converting reflected seismic waves into subsurface properties. The physics of wave propagation determines the relation between such reflected wavefields and the subsurface: this can vary from rather simple mathematical models that assume 1D layered subsurface structures to more advanced models that can handle complex, heterogeneous geologies. One such model relates pre-stack (or post-stack) seismic data to the elastic (or acoustic) reflectivity of the Earth: in other words,  primary only, P-wave seismic arrivals are described as the convolution of a source signature (i.e., wavelet) with the elastic (or acoustic) reflection response of the subsurface \cite{Goupillaud1961},
\begin{equation}
    d(t) = w(t) \ast r_{pp}(t),
\label{eq:seismic_data}
\end{equation}
where $r_{PP}$ is the P-wave reflectivity. Under the Fatti approximation \citep{Fatti1994} of the Zoeppritz equations \citep{Zoeppritz1919}, $r_{PP}$ is given by
\begin{equation}
    r_{PP}(t) = \sum_{i=1}^3 c_i(t, \theta)\frac{d}{dt}\log(m_i(t)),   
\label{eq:reflectivity}
\end{equation}
where the coefficients $c_i$ are given by
\begin{equation}
\begin{aligned}
    c_1(t, \theta) & = 0.5(1 + \tan^2(\theta))\\
    c_2(t, \theta) & = -4\left(\frac{\bar{V}_s(t)}{\bar{V}_p(t)}\right)^2\sin^2(\theta)\\
    c_3(t, \theta) & = 0.5\left(4\left(\frac{\bar{V}_s(t)}{\bar{V}_p(t)}\right)^2\sin^2(\theta) - \tan^2(\theta)\right),
\end{aligned}
\end{equation}
where $\theta$ is the angle of incidence, $\bar{V}_s$ is the shear wave velocity and $\bar{V}_p$ is primary wave velocity, and
\begin{equation}
    m(t) = (m_1(t), m_2(t), m_3(t)) = (AI(t), SI(t), \rho(t)),
\end{equation}
where $AI$ is the P-wave impedance, $SI$ is the S-wave impedance and $\rho$ is the density. In such a scenario, \eqref{eq:seismic_data} is usually referred to as the \textit{pre-stack seismic reflectivity}. When we consider only the zero-offset data, i.e. $\theta = 0$, the reflection response becomes
\begin{equation}
    r_{PP} = 0.5\frac{d}{dt}\log(AI(t)),
\label{eq:post-stack}
\end{equation}
known as the \textit{post-stack seismic reflectivity}. This is the physical model considered in the work, nevertheless all our findings can be also applied also to the physical model in equation \ref{eq:reflectivity}, as well as to other more complex modelling operators.

Post-stack seismic inversion refers to the process of retrieving an acoustic impedance model via equation \eqref{eq:seismic_data} where $r_{PP}$ is given by equation \eqref{eq:post-stack}. The associated modelling operator can be written in a compact matrix-vector notation as follows
\begin{equation}
    d = WDm,
\label{eq:post-stack-matrix}
\end{equation}
where $W$ represents the convolution operator with a given wavelet $w(t)$ (and is therefore a Toeplitz matrix), $D$ is a discretization operator of the first-order derivative, and $m = 0.5\log(AI(t))$. We define $G := WD$ to denote the post-stack seismic operator. Equation \eqref{eq:post-stack-matrix} is commonly solved for $m$ via the following minimization problem:
\begin{equation}
    \min_m \frac{1}{2}\Vert Gm - d \Vert_2^2.
\label{eq:post-stack-lsqr}
\end{equation}
Given the band-limited nature of the seismic wavelet, $W$ acts as a band-limiting operator, essentially masking the low and high frequencies of the sought-after reflectivity. Therefore, equation \eqref{eq:post-stack-matrix} is ill-posed, meaning that either no solution exists, the solution is not unique, or it is unstable with respect to small perturbations of the data. In order to obtain a meaningful solution, prior knowledge about the structure of $m$ has to be incorporated in the inversion. This information, commonly referred to as \textit{regularization}, is usually encoded via a norm and added as a penalty or a constraint to the objective function in equation \eqref{eq:post-stack-lsqr}, yielding
\begin{equation}
    \min_m \frac{1}{2}\Vert Gm - d \Vert_2^2 + \mathcal{R}(m).
\label{eq:post-stack-regularized}
\end{equation}
A classical choice for post-stack seismic inversion is $\mathcal{R}(x) = \lambda\Vert \nabla^2 x\Vert_2^2$ \citep{Tikhonov1977} or its directional form: this regularization enforces the solution to be smooth along the chosen direction(s), since it damps the amplitude of the second derivative. A more recent and appropriate choice is Total-Variation (TV) regularization \citep{Rudin1992}. Two versions of TV regularization exist: an anisotropic version, where $\mathcal{R}(x) = \lambda\Vert\nabla x\Vert_1$, and an isotropic version where $\mathcal{R}(x) = \lambda\Vert\nabla x\Vert_{21}$, with $\Vert A\Vert_{21} = \sum_{j=1}^n \Vert a_j\Vert_2$ denoting the matrix $2,1$-norm. TV regularization enforces the gradient to be sparse, thereby ensuring a solution that is mostly constant and has a few (possibly large) jumps. This behavior is particularly suitable for acoustic impedance models, making it a suitable choice for post-stack seismic inversion. 

\subsection{The primal-dual algorithm}
Proximal operators became popular in various imaging sciences because they allow efficient minimization of composite objective functions with one smooth and one or more non-smooth terms. These objective functions are also commonly used in various geophysical inverse problems: they consist of a smooth data term and a non-smooth regularization term, that is typically some form of sparsity promoting regularization. Let us write a generic objective function as 
\begin{equation}
    \mathcal{J}(x) = f(x) + g(Kx),
\end{equation}
where $f$ is the smooth data fidelity term, $g$ is the (possibly) non-smooth regularization term, and $K$ is a linear operator that transforms the solution into a domain where a suitable regularization can be applied. A prominent example of this type of functional is the state-of-the-art Total-Variation (TV) regularization
\begin{equation}
    \mathcal{J}(m) = \frac{1}{2}\Vert Gm - d\Vert_2^2 + \lambda\Vert \nabla m\Vert_1,
\label{eq:tvreg}
\end{equation}
where $f(m) = \frac{1}{2}\Vert Gm - b\Vert_2^2$, $g(m) = \lambda\Vert x\Vert_1$ and $K = \nabla$. The \textit{proximal operator} of a function $f$ is defined as 
\begin{equation}
    \text{prox}_f(v) = \argmin_x \left\{ \frac{1}{2}\Vert x - v\Vert_2^2 + f(x) \right\},
\label{eq:prox}
\end{equation}
which can be written in an alternative form by means of the optimality condition 
\begin{equation}
\begin{aligned}
    0 & \in x - v + \partial f(x) \\
    (I + \partial f)(x) & = v \\
    x & = (I + \partial f)^{-1}(v) \\
    \Longrightarrow \text{prox}_f(v) & = (I + \partial f)^{-1}(v)
\end{aligned}
\end{equation}
To deal with the non-smoothness of the function $g$, proximal algorithms introduce an auxiliary variable $y = Kx$ and solve
\begin{equation}
    \min_{x, y} f(x) + g(y) \quad \text{s.t.} \quad y = Kx.
\label{eq:Lagrange}
\end{equation}
There exist a number of algorithms that arise from solving equation \eqref{eq:Lagrange} in different ways. In this work, we focus on the primal-dual formulation \citep{Chambolle2011} since it has shown superior performance for TV-regularized post-stack seismic inversion \citep{Ravasi2022}. The Primal-Dual algorithm solves equation \eqref{eq:Lagrange} via the method of Lagrange multipliers. The Lagrangian yields
\begin{equation}
\begin{aligned}
    \max_{z}\min_{x, y} f(x) + g(y) + z^T(Kx - y) = \max_{z} \left(\min_{x} (f(x) + z^TKx) + \min_{y} (g(y) - z^Ty) \right)
\end{aligned}
\label{eq:lagrange}
\end{equation}
By using the definition of the \textit{conjugate function}
\begin{equation}
    f^{\star}(z) = \sup_x \{z^Tx - f(x)\} = -\inf_x \{ - z^Tx + f(x)\},
\end{equation}
and applying it to the function $g$ in equation \eqref{eq:lagrange}, we obtain 
\begin{equation}
    \max_z \min_x z^TKx + f(x) - g^{\star}(z).
\label{eq:primal-dual}
\end{equation}
If we further apply the definition of the conjugate function for $f$ to equation \ref{eq:primal-dual}, this yields the \textit{dual problem} 
\begin{equation}
    \max_z - f^{\star}(K^Tz) - g^{\star}(z).
\label{eq:dual}
\end{equation}
Starting from equation \ref{eq:primal-dual} we find the optimality conditions
\begin{equation}
\begin{aligned}
    0 & \in & \partial f(x) + K^Tz \\
    0 & \in & -\partial g^{\star}(z) + Kx
\end{aligned}
\end{equation}
Adding $x$ on both sides of the first equation and $z$ on both sides of the second equation yields
\begin{equation}
\begin{aligned}
    (I + \partial f)(x) & \in & x - K^Tz \\
    (I + \partial g^{\star})(z) & \in & z + Kx,
\end{aligned}
\end{equation}
which finally leads to
\begin{equation}
\begin{aligned}
    x & \in & (I + \partial f)^{-1}(x - K^Tz) = \text{prox}_f(x - K^Tz)\\
    z & \in & (I + \partial g^{\star})^{-1}(z + Kx) = \text{prox}_{g^{\star}}(z + Kx).
\end{aligned}
\end{equation}
Given this fixed-point condition, it is straightforward to derive the iterations of the primal-dual algorithm, also called the \textit{Chambolle-Pock algorithm} \citep{Chambolle2011},
\begin{equation}
\begin{aligned}
    x_{k+1} & = \text{prox}_f(\bar{x}_k - \sigma K^Tz_k) \\
    z_{k+1} & = \text{prox}_{g^{\star}}(z_k + \tau Kx_{k+1}) \\
    \bar{x}_{k+1} & = x_{k+1} + \theta(x_{k+1} - x_k),
\end{aligned}
\label{eq:CP}
\end{equation}
where the last step is a momentum step, added to accelerate convergence, and $\sigma$ and $\tau$ represent the step sizes applied to the two gradient terms. Note that the $z$ variable is updated using the proximal of the conjugate of $g$, which can be easily computed using the proximal operator of $g$ through the Moreau identity \citep{Parikh2014}:
\begin{equation}
    \text{prox}_{\tau g}(x) = x - \tau\text{prox}_{g^{\star}/\tau}(x/\tau)
\label{eq:Moreau}
\end{equation}
Typically, we choose $g = \mathcal{R}$ to be our regularization term, therefore the associated proximal operator can be written as 
\begin{equation}
    \text{prox}_{\sigma \mathcal{R}}(v) = \argmin_x \left\{ \frac{1}{2 \sigma}\Vert x - v\Vert_2^2 + \mathcal{R}(x) \right\}.
\label{eq:prox_denoise}
\end{equation}
The solution of this inverse problem can be interpreted as the MAP estimate of a denoising problem where the measurements $v $ are corrupted with zero-mean Gaussian noise with variance $\sigma$. Note that this interpretation of the proximal operator cannot be extended to the $f$ term, since $f$ is a data misfit term that does not correspond to any type of regularization, thereby losing the connection to a denoiser.

\subsection{The Plug-and-Play algorithm}
Given the interpretation of equation \eqref{eq:prox_denoise} as a denoising problem, \cite{Venkatakrishnan2013} suggest to replace such a step with any denoiser of choice. Starting from equation \ref{eq:CP}, it yields the following algorithm
\begin{equation}
\begin{aligned}
    x_{k+1} & = \text{prox}_f(\bar{x}_k - \sigma K^Tz_k) \\
    z_{k+1} & = \text{Denoise}(z + \tau Kx_{k+1}) \\
    \bar{x}_{k+1} & = x_{k+1} + \theta(x_{k+1} - x_k).
\end{aligned}
\label{eq:PnP}
\end{equation}
Note that in the seminal work of \cite{Venkatakrishnan2013}, the PnP algorithm was actually derived starting from the Alternating Direction Method of Multipliers (ADMM -- \citep{Boyd2011}); however, similar derivations with other proximal methods exist in the literature, and the one described in equation \ref{eq:PnP} was first proposed in \cite{Meinhardt2017}.

\subsection{Plug-and-Play as consensus equilibrium}
In general, one can write a regularized inverse problem as
\begin{equation}
    \min_x \sum_{i=1}^{n} f_i(x).
\end{equation}
For example, in case of $n = 2$ and 
\[
    f_1(m) = \frac{1}{2}\Vert Gm - d\Vert_2^2, \quad f_2(m) = \lambda\Vert \nabla m\Vert_1
\]
we obtain the previously defined TV-regularized linear inverse problem (equation \ref{eq:tvreg}). \textit{Consensus optimization} generalizes this concept to an arbitrary number of functions by defining the following objective function 
\begin{equation}
    \min_x \sum_{i=1}^{n} f_i(x_i) \text{ s.t. } x_i = x, \quad i=1,\ldots, n.
\label{eq:CO}
\end{equation}
where $n-1$ auxiliary variables are defined. Again, in case of $n = 2$, this reduces to
\begin{equation}
    \min_x f_1(x_1) + f_2(x_2) \text{ s.t. } x_1 = x, x_2 = x,
\end{equation}
which is equivalent to the variable splitting approach in equation \ref{eq:Lagrange} with $x_1=x$, $x_2=y$, $K=I$, $f_1=f$, and $f_2=g$. This general formulation can be extended even further by imposing \textit{consensus equilibrium}, where
\begin{equation}
\begin{aligned}
    F_i(x^{\star} + u_i^{\star}) & = x^{\star} \\
    \bar{u}^{\star}_{\mu} & = 0,
\end{aligned}
\label{eq:CE}
\end{equation}
where $\bar{u}_{\mu} = \sum_{i=1}^n \mu_iu_i$ with $\sum_{i=1}^n \mu_i = 1$. \cite{Buzzard2018} showed, that if
\begin{equation}
    F_i(x) = \argmin_v \left\{ \frac{1}{2\sigma} \Vert x-v\Vert_2^2 + f_i(v)\right\},   
\end{equation}
then equations \eqref{eq:CO} and \eqref{eq:CE} are equivalent.

\subsection{Multi-Agent Consensus Equilibrium (MACE)}
Multi-Agent Consensus Equilibrium (MACE) is an optimization algorithm that aims to enforce consensus equilibrium. Here, the various functions $F_i$ are referred to as \textit{agents}. The benefit of MACE is that it assumes no underlying objective function, thereby allowing more freedom in selecting such agents. For example, certain denoisers, like BM3D, do not have an underlying objective function that is being solved, but they could be easily incorporated as an agent into MACE. As such, MACE can solve inverse problems by including priors that are not a norm alongside some other explicit functions of the underlying variable, such as the $\ell_2$-norm and $\ell_1$-norm. Let us define the following operators
\begin{equation}
    \mathbf{F}(\mathbf{x}) = \left[ F_1(x_1)\: ; \ldots; F_n(x_n)\right], \quad \mathbf{K}(\mathbf{x}) = \left[ \bar{x}\: ; \ldots; \bar{x}\right], \quad  \bar{x} = \frac{1}{n}\sum_{i=1}^n x_i.
\end{equation}
where $\mathbf{x} = \left[ x_1\: ; \ldots; x_n\right]$. MACE achieves consensus equilibrium by finding a solution $\mathbf{x}^*$ that satisfies
\begin{equation}
    \mathbf{F}(\mathbf{x}^*) = \mathbf{K}(\mathbf{x}^*). 
\label{eq:FK}
\end{equation}
Although Newton's method could be applied to solve such an equality via
\[
    \mathbf{F}(\mathbf{x}) - \mathbf{K}(\mathbf{x}) = 0,
\]
this requires computing the gradients of each function $F_i$, which potentially might not exist. An alternative way to find the solution $\mathbf{x}^*$ is to use fixed-point iterations. First, note that equation \eqref{eq:FK} is equal to 
\begin{equation}
    (2\mathbf{F} - I)(\textbf{x}^*) = (2\mathbf{K} - I)(\textbf{x}^*).
\label{eq:2FI}
\end{equation}
Since $\mathbf{K}^2 = \mathbf{K}$, we have
\[
    (2\mathbf{K} - I)(2\mathbf{K} - I) = 4\mathbf{K}^2 - 4\mathbf{K} + I = I.
\]
Multiplying each side equation \eqref{eq:2FI} by $(2\mathbf{K} - I)$ yields
\begin{equation}
    (2\mathbf{K} - I)(2\mathbf{F} - I)(\textbf{x}^*) = \textbf{x}^*,
\label{eq:2KFI}
\end{equation}
which justifies the fixed-point iteration
\[
    \textbf{x}_{k+1} = T(\textbf{x}_k) := (2\mathbf{K} - I)(2\mathbf{F} - I)(\textbf{x}_k),
\label{eq:fixed-point}
\]
where $T$ is the fixed-point operator. Convergence of the fixed-point iteration is only guaranteed if $T$ is non-expansive, which for a linear operator means that the largest eigenvalue of $T$ is smaller than 1. It is, however hard to evaluate whether the consensus equilibrium operator $T$ is non-expansive given the presence of possibly non-linear $F_i$ agents; luckily, one can resort to the \textit{Mann iteration}, which is guaranteed to be non-expansive for any operator $T$,
\begin{equation}
    \textbf{x}_{k+1} = \left(\frac{1}{2}I + \frac{1}{2}T\right)(\textbf{x}_k)
\label{eq:Mann}
\end{equation}
Finally, without loss of generality, we may choose to not distribute the weights uniformly in $\mathbf{\bar{x}}$, and define
\begin{equation}
    \mathbf{\bar{x}} = \sum_{i=1}^n w_ix_i, \quad \sum_{i=1}^n w_i = 1,  
\label{eq:weighted_sum}
\end{equation}
so that we can attribute more importance to certain agents than others.

\subsection{Extending Plug-and-Play to 3D using MACE}
The superiority of PnP over traditional regularization methods such as TV-regularization has been previously shown in a variety of inverse problems, including post-stack seismic inversion \citep{Romero2023c}. However, although extending traditional regularizers from 2D to 3D is trivial, a difficulty with PnP priors lies in the fact that while there is an abundance of pre-trained 2D denoisers, pre-trained 3D denoisers are not widely available and are more expensive to train and apply. Using the MACE framework, we can introduce a sequence of denoisers that denoise 2D slices of a 3D model that we wish to estimate along all possible directions of the model, i.e. in the inline direction, the crossline direction, and the time direction. Combining them with the post-stack seismic inversion operator allows for 3D post-stack seismic inversion to be performed with 2D pre-trained denoisers. In mathematical terms, we define 
\[
    F_1(v) = \text{prox}_f(v) = \argmin_x \left\{\frac{1}{2}\Vert x - v\Vert_2^2 + \frac{1}{2}\Vert Gx - d\Vert_2^2\right\},
\]
where $G$ is the post-stack seismic modeling operator. Given $X \in \mathbb{R}^{n_{il}\times n_{xl}\times n_{t}}$ (where $x=Vec(X)$), we further define 
\begin{equation}
\begin{aligned}
    F_2(X)[i] & = \text{Denoise}(X[i]) \\
    F_3(X)[:,i] & = \text{Denoise}(X[:,i]) \\
    F_4(X)[:,:,i] & = \text{Denoise}(X[:,:,i]).
\end{aligned}
\label{eq:3D_denoisers}
\end{equation}
where $i$ indicates that each slice of the model, taken in any of the chosen directions, is denoised independently from the others. Consensus equilibrium now requires that the output of all denoisers and the result of the proximal operator $F_1$ are the same, meaning that denoising applied along the three directions of the seismic cube gives a consistent result.

\subsection{Choice of denoiser}
Two families of denoisers exist in the literature, namely blind and non-blind denoisers. Blind denoisers do not take the noise level of the data into account, whereas non-blind denoisers require an extra parameter as input that represents the noise level itself. Clearly, the proximal operator in equation \eqref{eq:prox} assumes knowledge of the noise level (used to scale the data misfit term). Therefore, it seems natural to replace such a proximal operator with a non-blind denoiser. Moreover, most pre-trained denoisers do not generalize well to datasets with different noise levels. For this purpose, \cite{Zhang2021} proposed the DRUNet, whose architecture is the combination of a UNet with a ResNet, and takes the noise level $\sigma$ as an additional input channel, and minimizes the following loss function:
\begin{equation}
    \mathcal{L}(\theta) = \sum_{i=1}^n \Vert\text{DRUNet}(y_i, \sigma_i;\theta) - x_i\Vert_1,
\end{equation}
where $\theta$ are the network parameters, $x_i$ are the clean images, $y_i$ are the noisy images, and $\sigma_i$ is the corresponding noise level. In this work, we rely on their pre-trained network and apply it directly to acoustic impedance models within the PnP and MACE iterations, without requiring any fine-tuning to the network itself.

\section{Results}

\subsection{2D post-stack seismic inversion on the SEAM data}
\begin{figure}
    \centering
    \begin{minipage}[t]{\linewidth} 
        \centering
        \includegraphics[width=\linewidth]{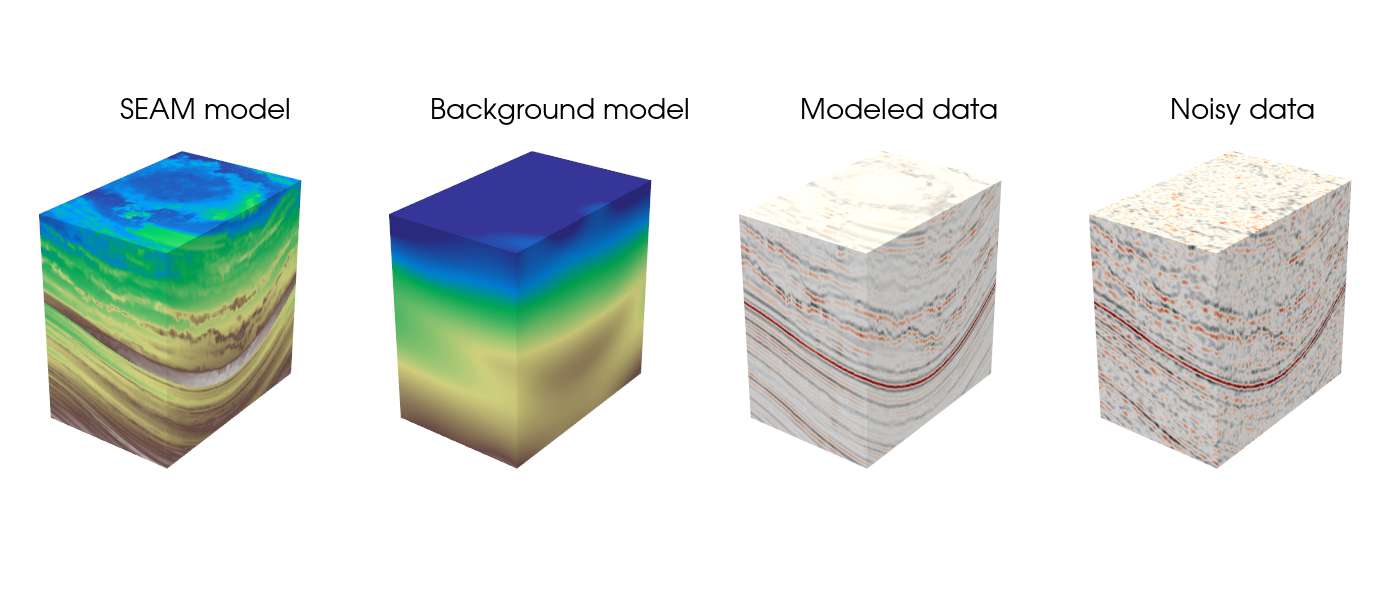} 
        \caption*{(a) Part of the 3D SEAM model, the background model, the corresponding modelled data and the noisy data.}
    \end{minipage}\hfill
    \begin{minipage}[t]{\linewidth} 
        \centering
        \includegraphics[width=\linewidth]{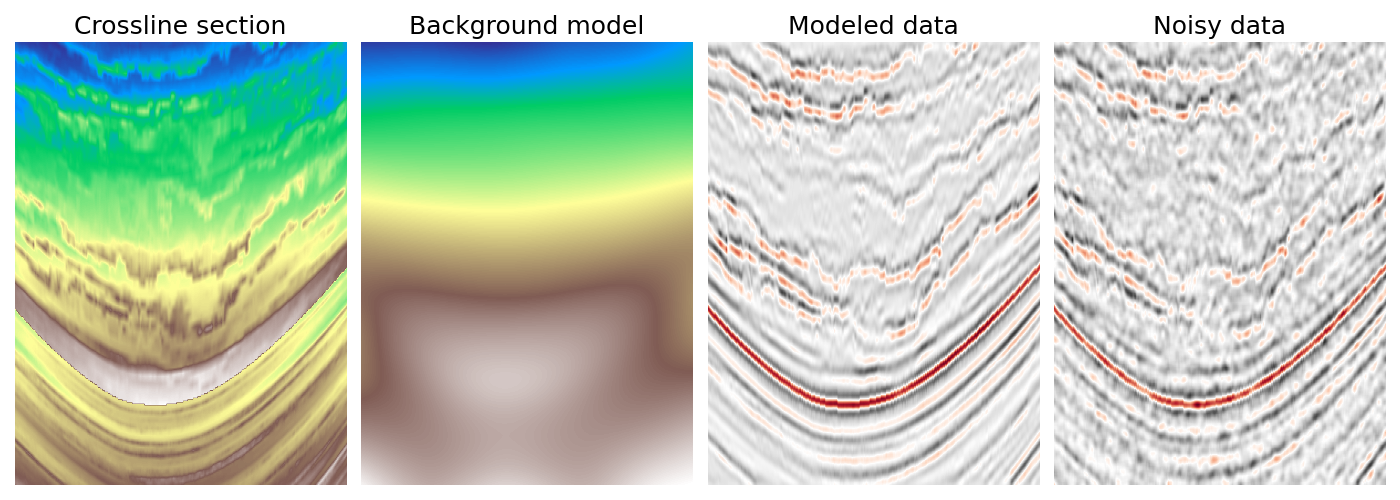} 
        \caption*{(b) Crossline section of the 3D model, background model, the corresponding modelled data and the noisy data for 2D inversion.}
    \end{minipage}
    \caption{}
    \label{fig:models}
\end{figure}

We start by making a comparison between TV-regularized post-stack inversion solved using the primal-dual method (TV-PD), PnP, and MACE. For all experiments, we use the SEAM Phase 1 acoustic velocity model\footnote{\url{https://seg.org/SEAM/open-data}}, where the 2D model is obtained by taking a slice in the middle of the crossline direction of the 3D model. Both the 3D and 2D models used in this work are shown in figure \ref{fig:models} (first panel in both a and b parts) alongside their background models (second panel).
The corresponding post-stack seismic datasets, modeled using the PyLops toolbox \citep{ravasi2020}, are shown in the third panel of figure \ref{fig:models}, and the noisy data obtained by adding colored Gaussian noise with standard deviation $\sigma = 0.2$ are shown in the last panel. 

\begin{figure}
    \centering
    \includegraphics[width=\linewidth]{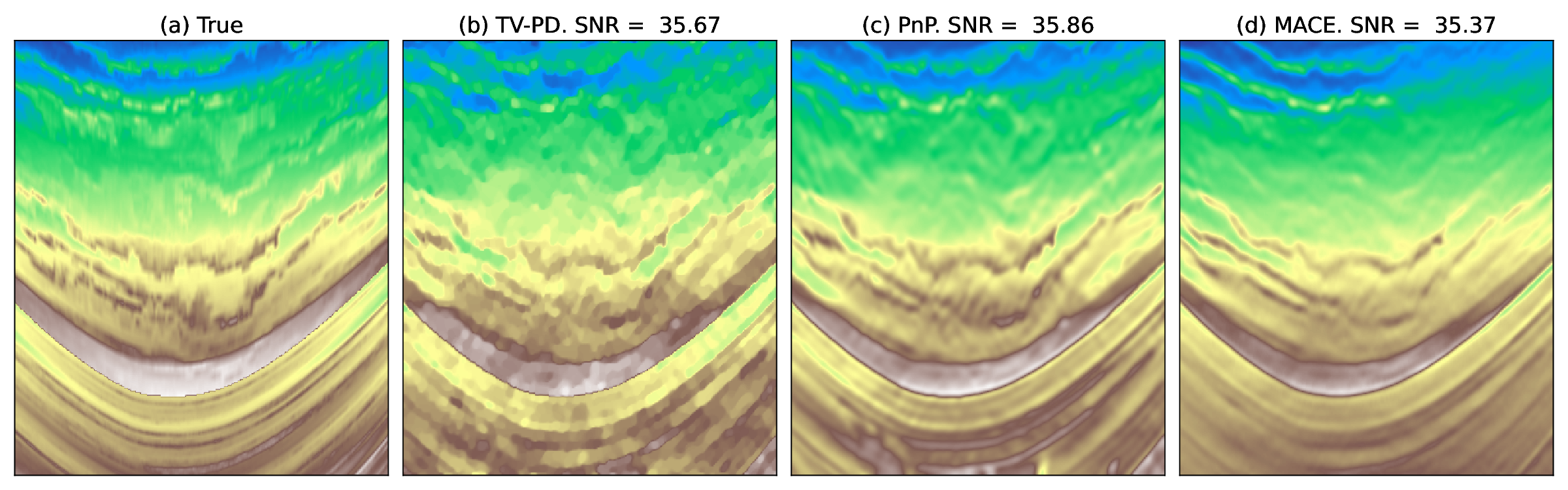} 
    \caption{Comparison of TV-PD, PnP and MACE for 2D post-stack seismic inversion.}
    \label{fig:2D_results}
\end{figure}

Figure \ref{fig:2D_results} shows the 2D inversion results for the 3 methods under investigation alongside the ground truth model.
Clear differences can be observed between the model obtained using the TV-PD algorithm and those from PnP and MACE. The TV-PD solution looks noisier, although some of the fine-scale details seem to be better retrieved and the magnitude of the high impedance layer better resolved. On the other hand, PnP and MACE solutions are similar, with the MACE solution being a bit over-smoothed. Note that equations \eqref{eq:CE} and \eqref{eq:CO} are expected to produce equivalent solutions only when all the agents $F_i$ are the proximal operators of the associated functions $f_i$. However, when one of the $F_i$ is a denoiser and one of the updates of the PD algorithm is replaced by a denoiser (meaning that an underlying objective function of the form \eqref{eq:CO} does no longer exist), this equivalency no longer necessarily holds true. Interestingly, although the TV-PD and PnP solutions look structurally very different, the Signal to Noise Ratio (SNR) are close, indicating that the SNR alone is not a representative measure of the quality of the reconstruction. In fact, when comparing the different solutions, it is clear that MACE and, to a lesser extent PnP, recover more of the high-resolution details in the middle of the model than TV-PD.

\begin{figure}
    \centering
    \includegraphics[width=\linewidth]{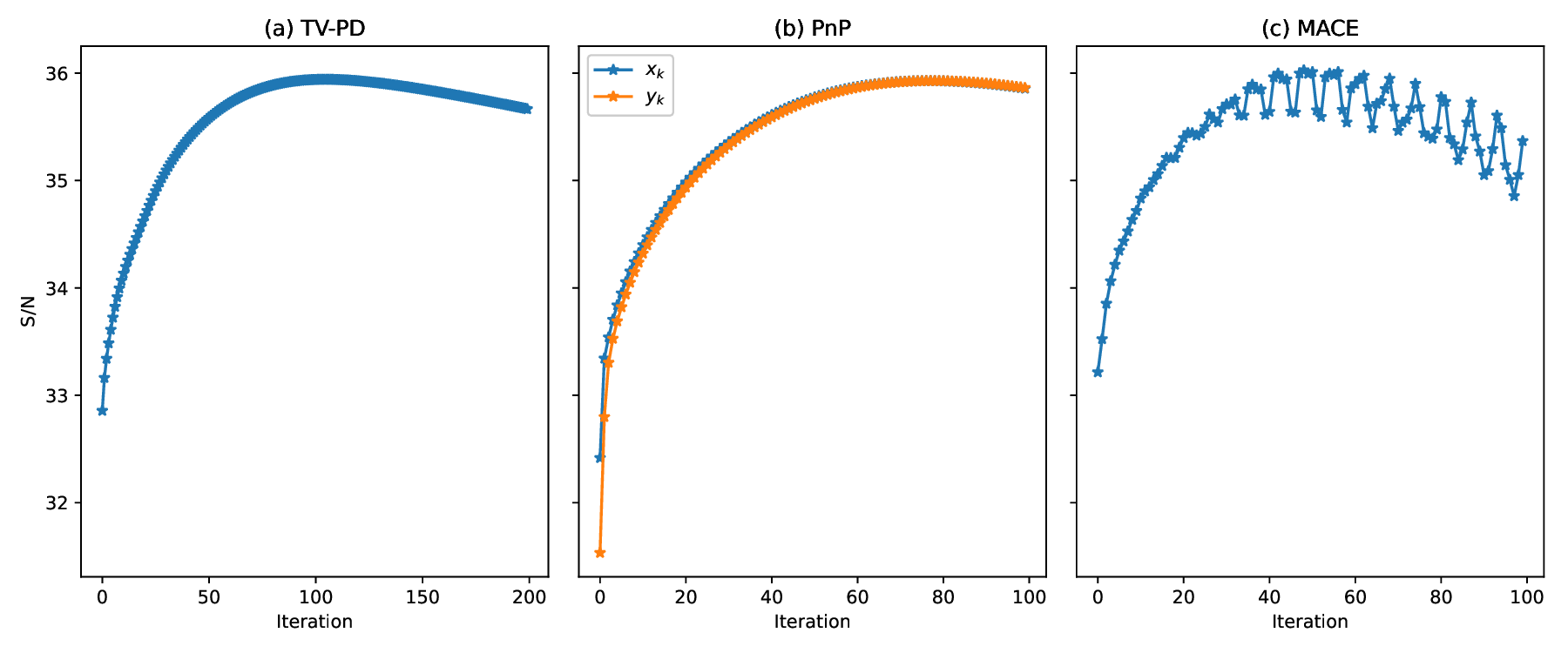} 
    \caption{Comparison of the convergence behaviour of TV-PD, PnP and MACE in terms of SNR for the 2D post-stack seismic inversion experiment.}
    \label{fig:convergence_2D}
\end{figure}

Finally, figure \ref{fig:convergence_2D} shows the SNR as a function of iterations. The SNR of the PnP method is the most stable over iterations with TV-PD being a close second, while the SNR of MACE fluctuates erratically after roughly 40 iterations. We note that, in practical applications, stable convergence is important since we usually rely on early stopping, and large variations in SNR between iterations may lead to inadvertently getting a poor result. 

\subsection{3D post-stack seismic inversion on the SEAM data}
\begin{figure}
    \centering
    \includegraphics[width=\linewidth]{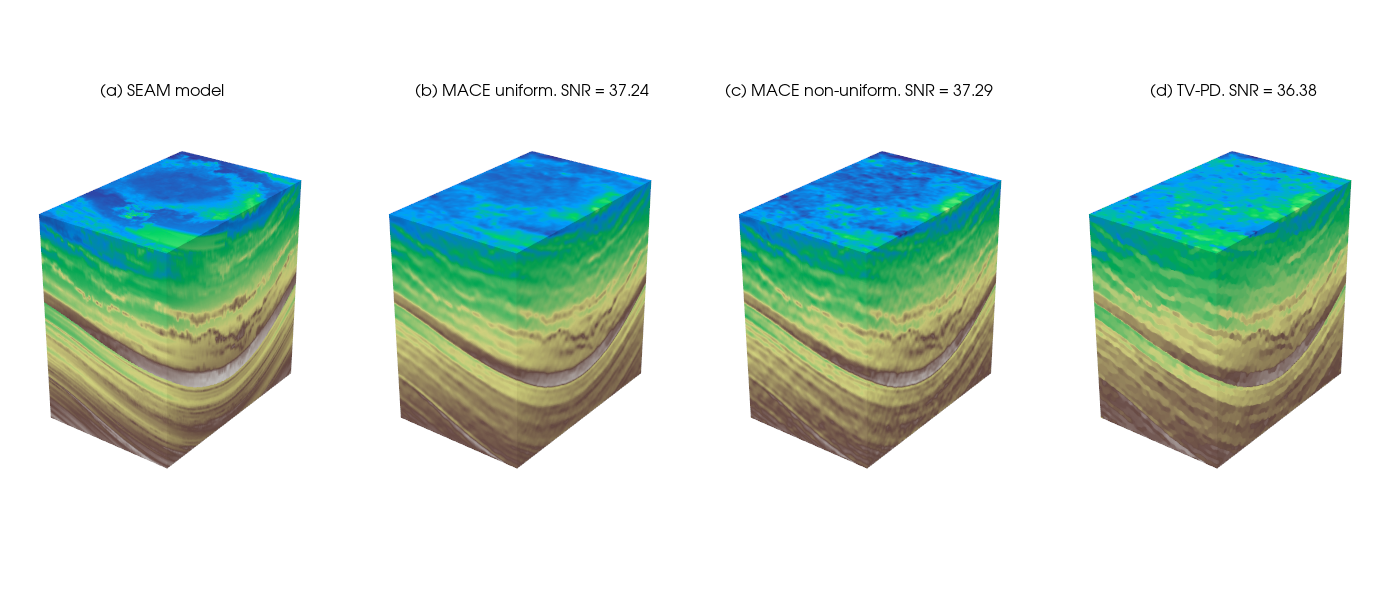} 
    \caption{Comparison of uniform and non-uniform MACE solutions.}
    \label{fig:3D_result}
\end{figure}

\begin{figure}
    \centering
    \begin{minipage}[t]{0.48\linewidth} 
        \centering
        \includegraphics[width=\linewidth]{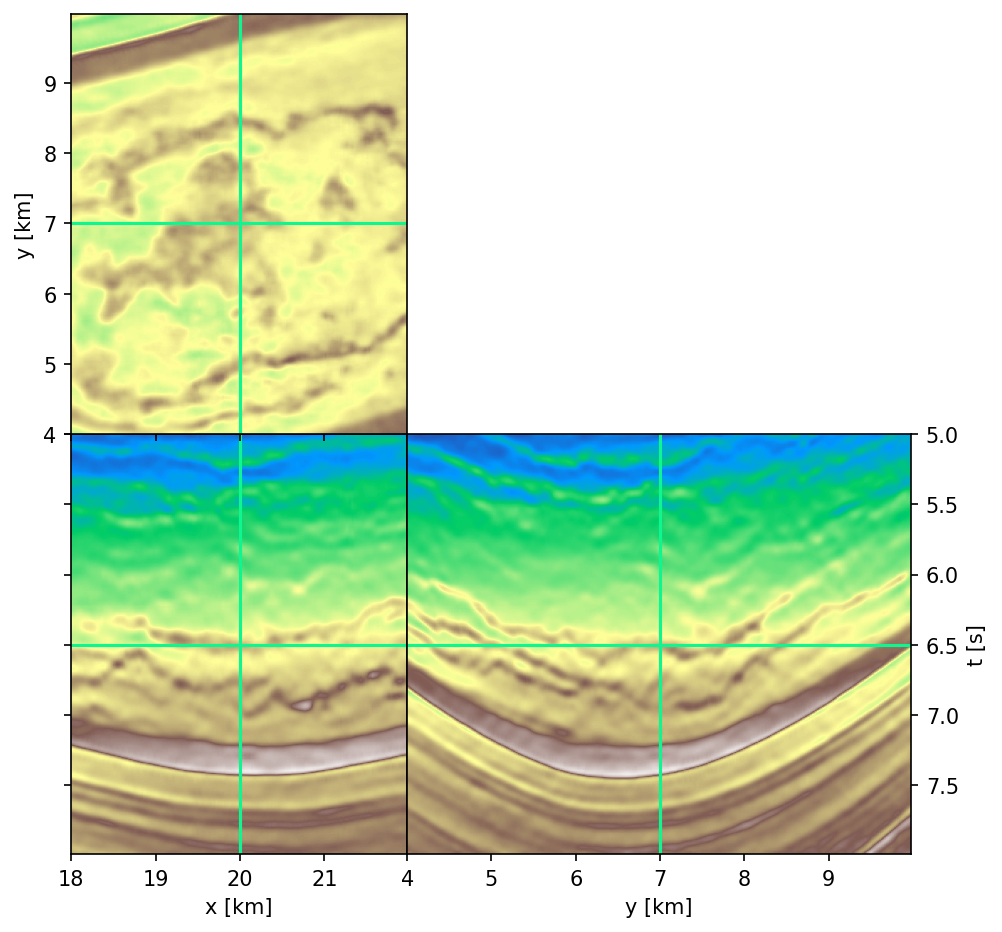} 
        \caption*{(a) Unfolded view of the MACE solution for the acoustic impedance model in figure \eqref{fig:3D_result}b. The green lines indicate the sections of the cube displayed in the different panels.}
    \end{minipage}\hfill
    \begin{minipage}[t]{0.48\linewidth} 
        \centering
        \includegraphics[width=\linewidth]{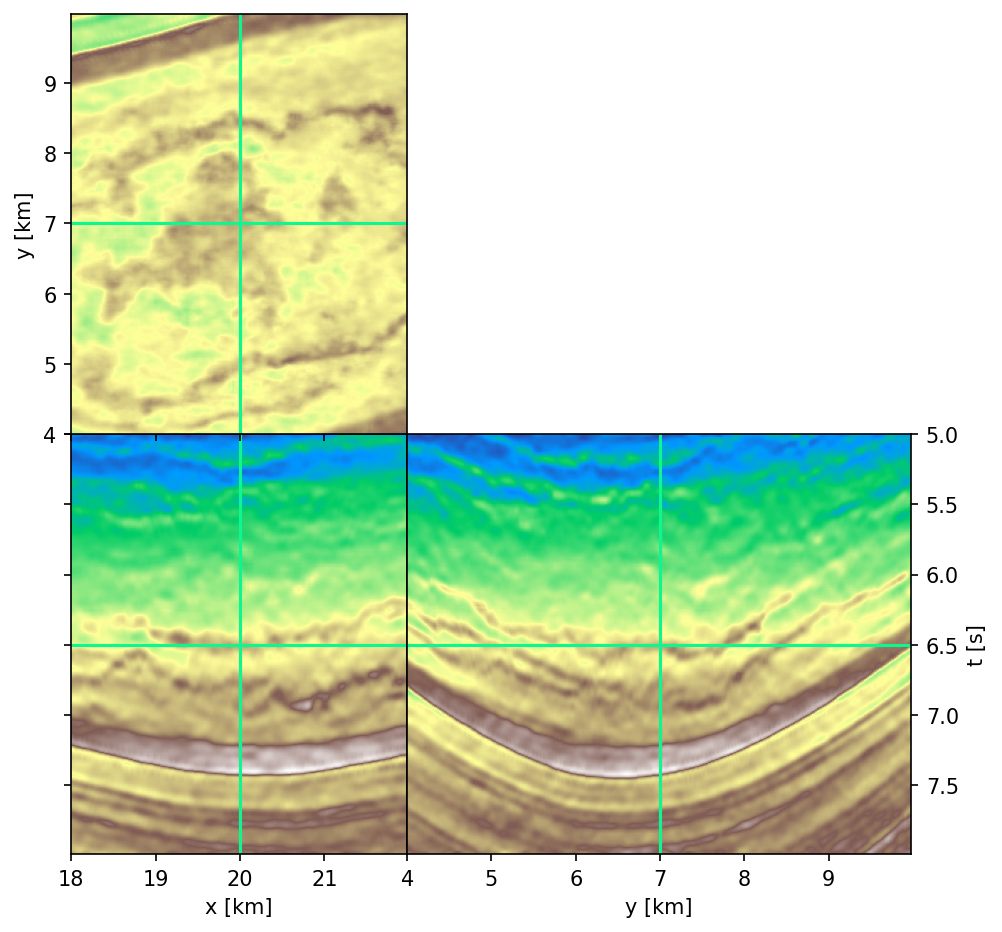} 
        \caption*{(b) Same as above for the acoustic impedance model in figure \eqref{fig:3D_result}c.}
    \end{minipage}
    \caption{}
    \label{fig:explode_result}
\end{figure}

We now consider a portion of the SEAM Phase 1 velocity model and perform 3D seismic inversion. Similar to the previous example, we compare here TV-PD and MACE with three denoising agents as described in equation \eqref{eq:3D_denoisers}. We consider two variants for the MACE algorithm that use different weights according to equation \eqref{eq:weighted_sum}: the first uses a uniform weight distribution $w_i = 1/4, i=1\ldots 4$, whilst the second uses $w_1 = 1/2$, $w_i = 1/6, i=2\ldots 4$, hereafter referred to as the solution with a non-uniform weight distribution. The second solution divides the weight between all denoisers and the physics equally, unlike the solution with a uniform weight distribution, which gives equal importance to all agents. The reconstructed 3D models are shown in figure \eqref{fig:3D_result}, and a 2D view of the unfolded 3D models is displayed in figure \eqref{fig:explode_result}.
Both the uniform and non-uniform solutions look better than the TV-PD solution, which presents some visible noise leakage from the migrated data. This is especially the case for some of the main layers, which are better captured by the two MACE solutions. In this case the SNR measure seem to corroborate our visual assessment.

\begin{figure}
    \centering
    \begin{minipage}[t]{0.5\linewidth} 
        \centering
        \includegraphics[width=\linewidth]{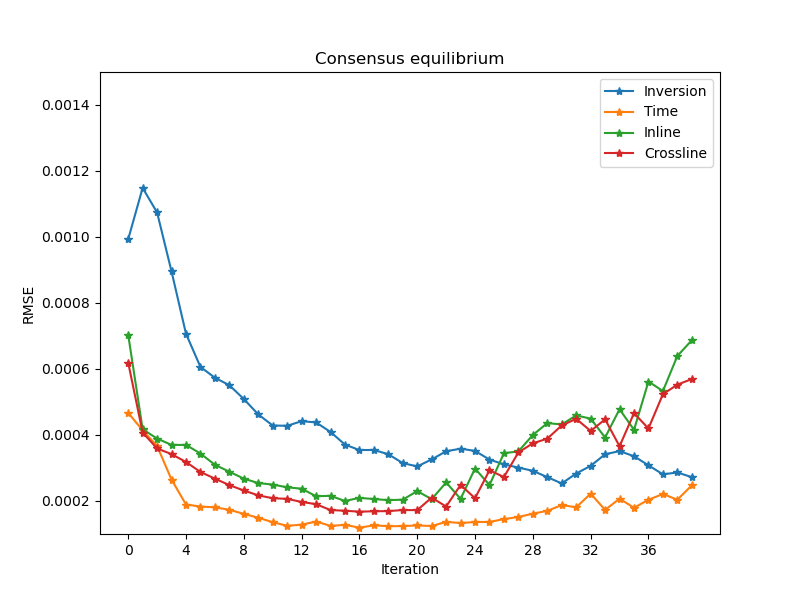} 
        \caption*{(a) Output of equation \eqref{eq:consensus_metric} for the uniform solution.}
    \end{minipage}\hfill
    \begin{minipage}[t]{0.5\linewidth} 
        \centering
        \includegraphics[width=\linewidth]{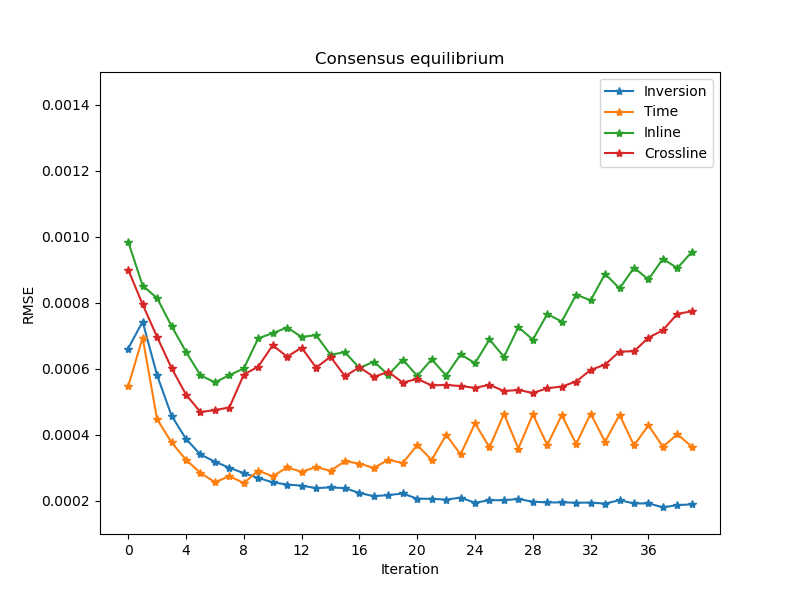} 
        \caption*{(b) Output of equation \eqref{eq:consensus_metric} for the non-uniform solution.}
    \end{minipage}
    \caption{}
    \label{fig:equilibrium}
\end{figure}

\begin{figure}
    \centering
    \begin{minipage}[t]{0.5\linewidth} 
        \centering
        \includegraphics[width=\linewidth]{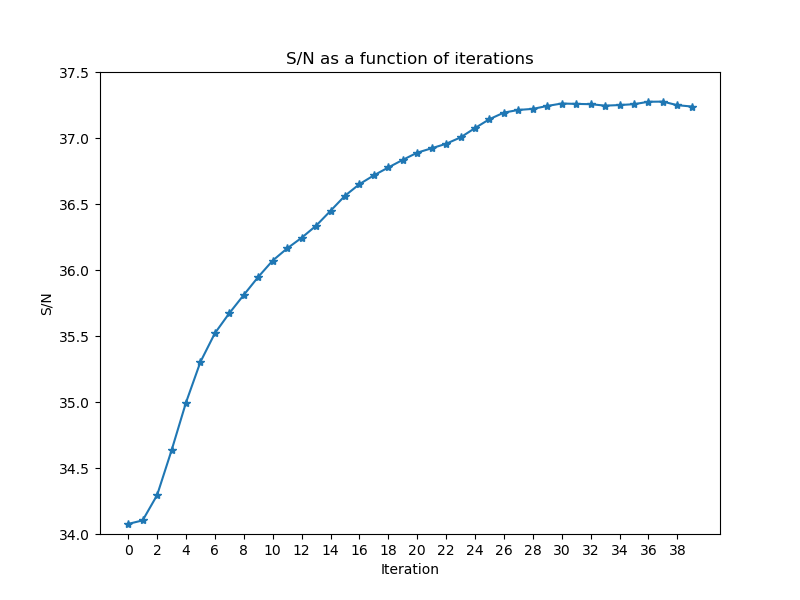} 
        \caption*{(a) SNR for the uniform solution.}
    \end{minipage}\hfill
    \begin{minipage}[t]{0.5\linewidth} 
        \centering
        \includegraphics[width=\linewidth]{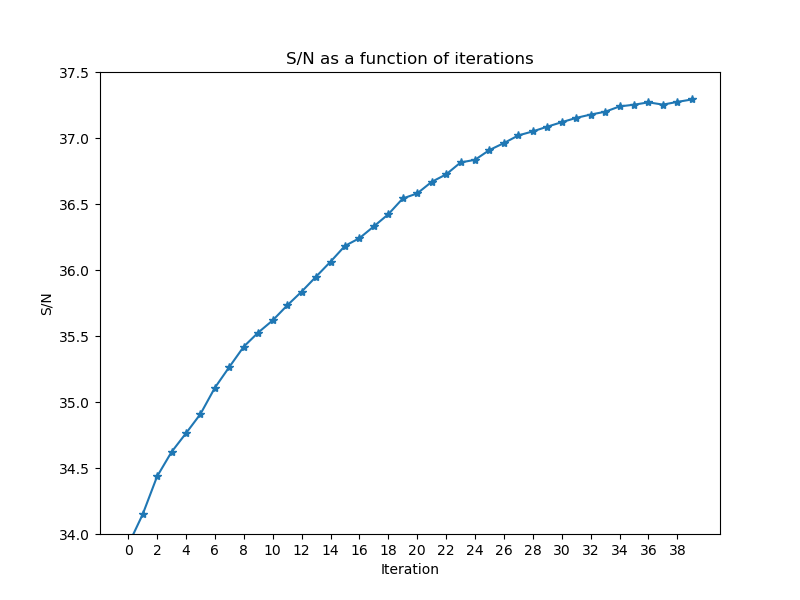} 
        \caption*{(b) SNR for the non-uniform solution.}
    \end{minipage}
    \caption{}
    \label{fig:S/N}
\end{figure}

\begin{figure}
    \centering
    \includegraphics[width=\linewidth]{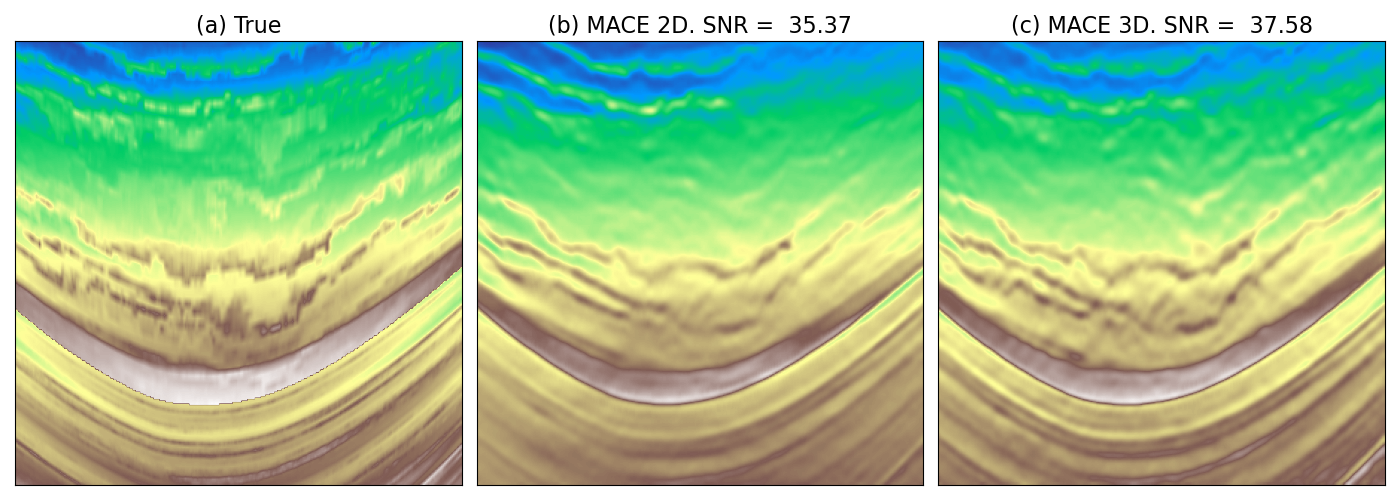}
    \caption{Comparison of the 2D result with the corresponding section of the 3D result.}
    \label{fig:2Dvs3D}
\end{figure}

Finally, when comparing the MACE solutions, we observe that the non-uniform solution is noisier than its uniform counterpart; however, the corresponding SNR is slightly higher. A look at figure \eqref{fig:explode_result} confirms this, although in this figure, the uniform MACE solution looks better due to the better continuity of some the main layers and the higher resolution of the syncline structure. This confirms once again that SNR alone is not a good enough indicator to favor one solution over another. In both experiments, MACE is run for a total of 40 iterations. In order to assess whether the algorithm has reached convergence, we track consensus equilibrium over iterations by computing the following metric
\begin{equation}
    \frac{\Vert\bar{x} - x_i\Vert_2^2}{\Vert\bar{x}\Vert_2^2 }, \quad i=1,\ldots, 4,
\label{eq:consensus_metric}
\end{equation}
for every output of the agents at every iteration. Consensus equilibrium is reached when this metric converges to zero. Figure \eqref{fig:equilibrium} shows the curves computed using equation \eqref{eq:consensus_metric} for both the uniform and the non-uniform MACE solutions.
For neither the uniform nor the non-uniform cases, consensus equilibrium is reached and the solutions seem to diverge. For the uniform solution, this seems to happen at around iteration 30, whilst for the non-uniform solution this happens already after 5 iterations. We further plot the corresponding SNR in figure \eqref{fig:S/N}, and notice that, interestingly a lack of consensus equilibrium is not reflected in the SNR of the solution. Ideally, when the agents start to diverge we would expect a decrease in SNR; however, this is not the case in our numerical example. This means that consensus equilibrium may not be a good indicator of the quality of the solution and different automated stopping criteria must be identified.
Lastly, we compare the 2D solution to the corresponding slice of the 3D solution in figure \ref{fig:2Dvs3D}. Clearly, the 3D result is of greater quality, indicating that MACE is able to exploit information from the 3D nature of the seismic data despite only applying 2D denoisers.

\subsection{3D post-stack seismic inversion on the Sleipner data}
\begin{figure}
    \centering
    \includegraphics[width=\linewidth]{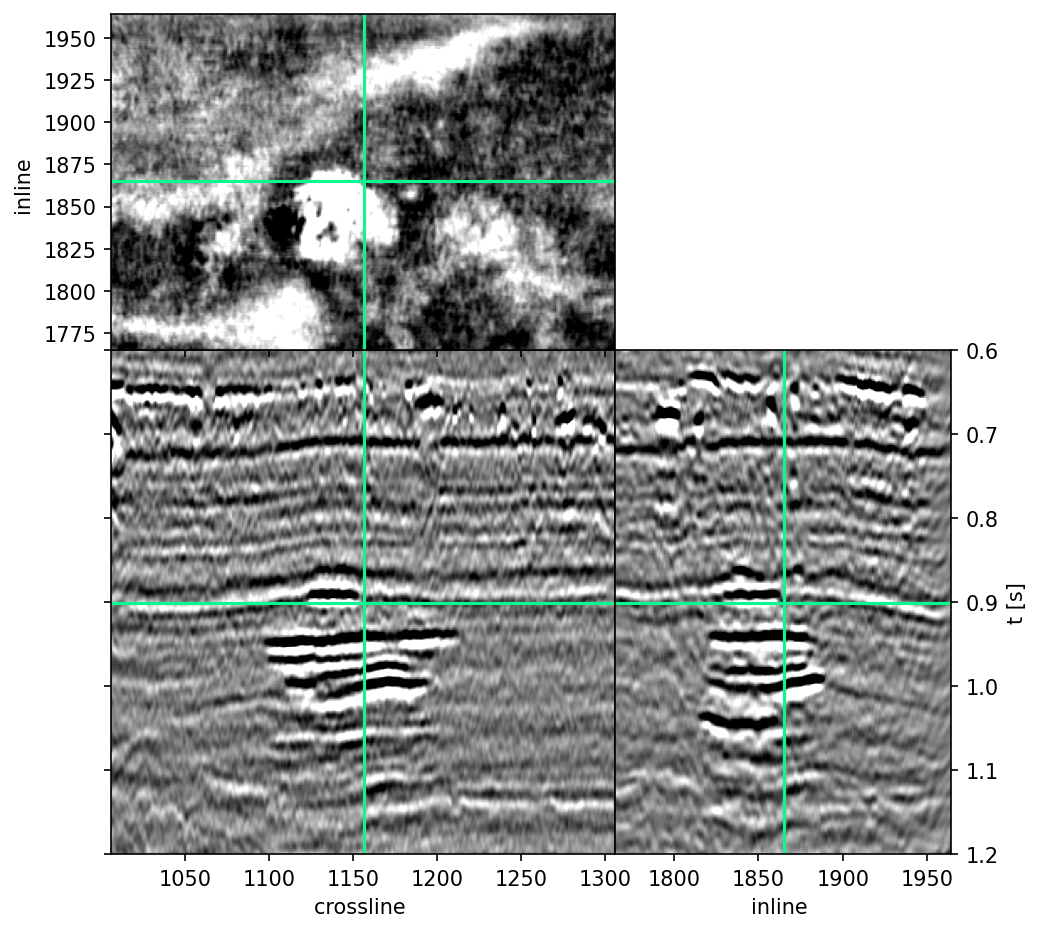}
    \caption{Sleipner data.}
    \label{fig:Sleipner_data}
\end{figure}

\begin{figure}
    \centering
    \begin{minipage}[t]{0.48\linewidth} 
        \centering
        \includegraphics[width=\linewidth]{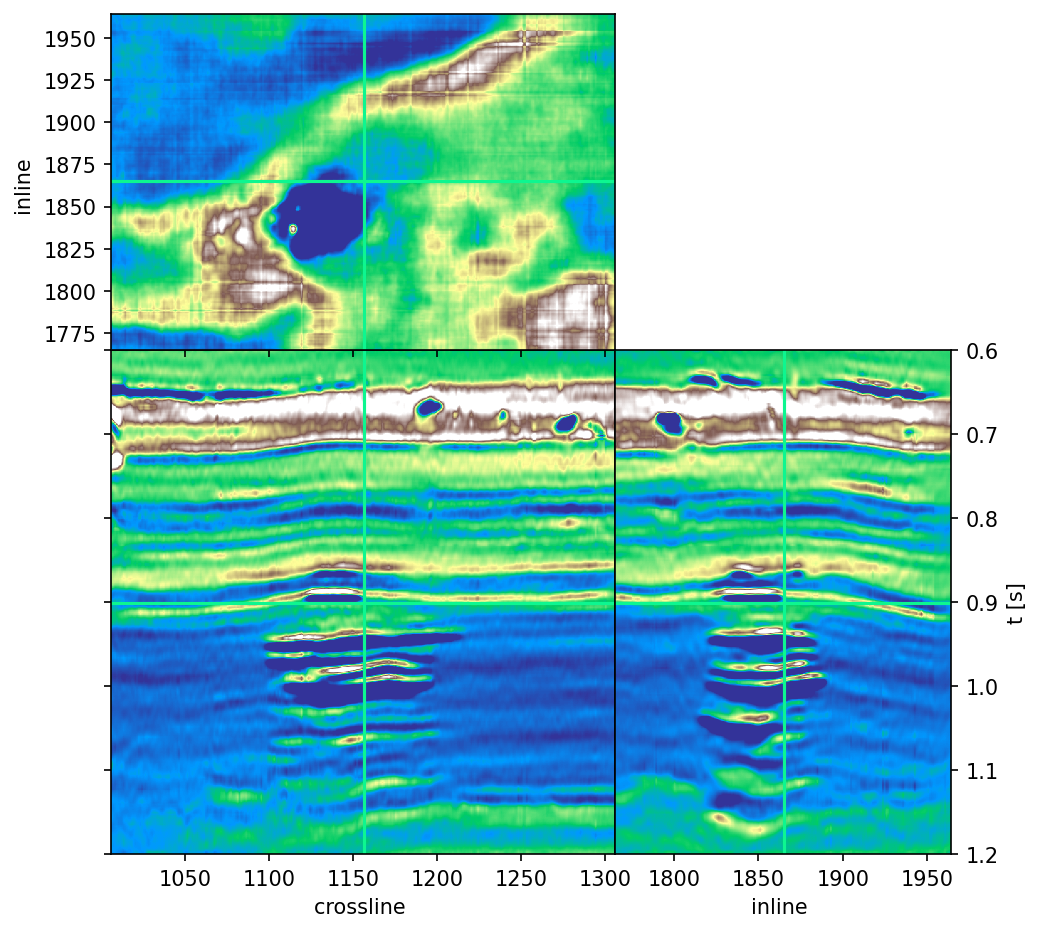} 
        \caption*{(a) Inverted acoustic impedance model using MACE.}
    \end{minipage}\hfill
    \begin{minipage}[t]{0.48\linewidth} 
        \centering
        \includegraphics[width=\linewidth]{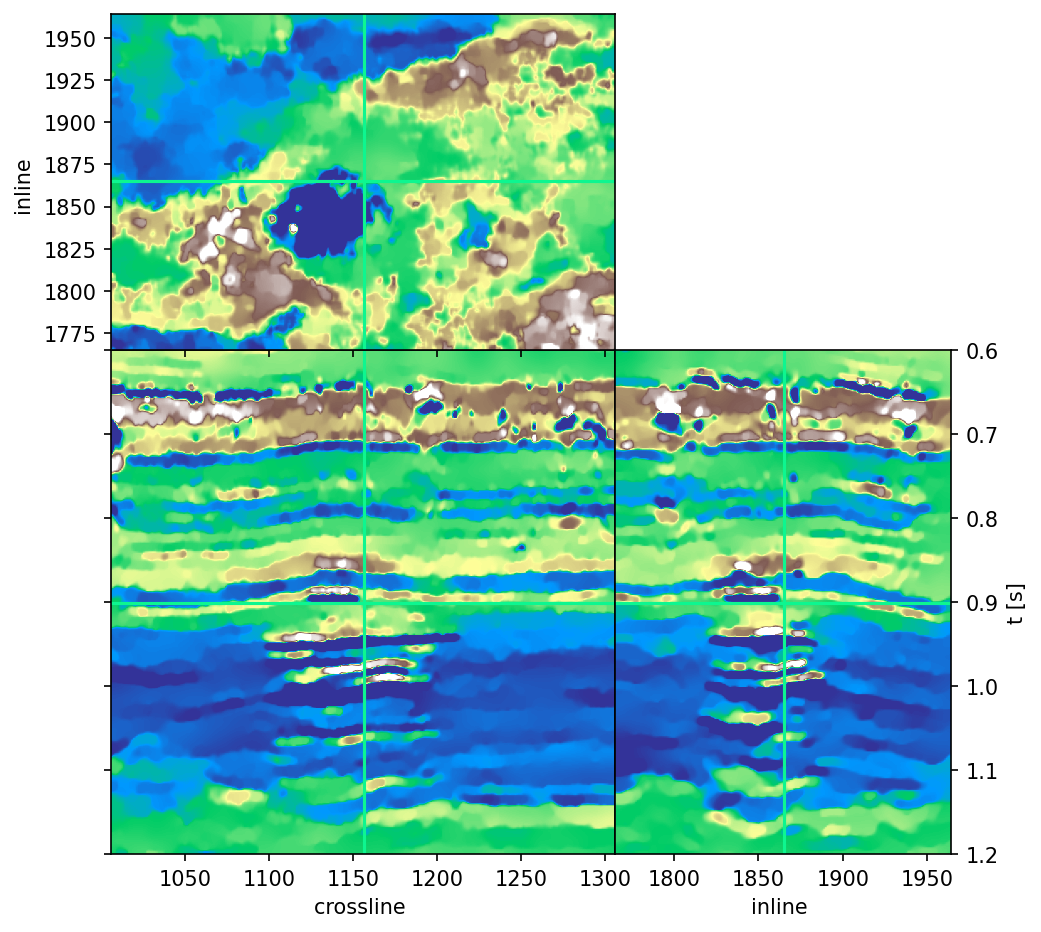} 
        \caption*{(b) Inverted acoustic impedance model using TV.}
    \end{minipage}
    \caption{}
    \label{fig:Sleipner_result}
\end{figure}

Finally, we test MACE on the Sleipner field dataset \citep{chadwick2010} and compare the results to TV-regularized inversion. The Sleipner dataset is a 4D seismic dataset consisting of 6 different seismic surveys, with a baseline acquisition in 1994, followed by 5 different monitor acquisitions in 1999, 2001, 2004, 2006, 2008 and 2010, with the goal of monitoring the injection of a $CO_2$ into the Utsira formation. In this work, we apply post-stack inversion to the 2001 monitor dataset. The data to be inverted is shown in figure \eqref{fig:Sleipner_data} and the inverted acoustic impedance models from MACE and TV-PD are shown in figures \ref{fig:Sleipner_result}a and \ref{fig:Sleipner_result}b. 
As discussed in more detail in \cite{Romero2023a, Romero2023b}, we first perform a well-to-seismic tie with well 15/9-13, after which a time-depth relationship is created using well picks and time reflectors from the baseline seismic volume. A smoothed version of the acoustic impedance depth profile is created from the density and P-wave velocity well logs and used to create a smooth, horizontally homogeneous background model. Statistical wavelet estimation is performed within a subvolume around the reservoir depth, and the estimated wavelet is subsequently scaled by comparing the modeled synthetic trace at the well location with the closest trace in the seismic data.

Similar to the results obtained in the previous example, the MACE solution for the Sleipner data looks of higher quality than its TV-PD counterpart. For example, some of the key layers are far more continuous in the MACE solution; this is especially the case for the high-impedance layer at the top of the model. Looking at the $C0_2$ plume, we observe similar characteristics: it is far easier to distinguish some of the thin layers in the MACE solution than in the TV-PD solution. However, when looking at a time slice, we notice some stripes along the inline and crossline directions for the MACE solution: this is currently interpreted as an imprint of the denoising process that is being applied for every slice individually.

\section{Discussion}
Our numerical experiments indicate that PnP and MACE can outperform traditional regularization methods such as TV regularization. One of the limiting factors of PnP is that the replacement of one of the proximal operators with a denoiser leads to a decoupling of the chosen optimization algorithm with the underlying objective function. \cite{Ryu2019} have laid the theoretical groundwork for PnP, proving convergence under certain conditions. However, some of these conditions are quite restrictive: for example, the data misfit must be strongly convex, which for linear problems with an $\ell_2$ data fidelity corresponds to having a full rank operator, a condition which is rarely met in practice, and certainly does not hold for post-stack seismic modelling operator. Moreover, the neural network-based denoiser needs to be Lipschitz continuous. Although this property can be enforced either in the design of the network or during training, this makes training much harder and potentially limits the expressiveness of the network. MACE takes a different starting point than PnP, by no longer requiring an underlying objective function: on the other hand, the method simply relies on agents that satisfy consensus equilibrium. Although there is a relation between consensus equilibrium and consensus optimization, this requires all agents to be proximal operators of some explicitly defined functions of the objective function in consensus optimization; this is of no practical in our case since denoisers are used in spite of proximal operators. Although MACE should reach consensus equilibrium after a given number of iterations,  this is not the case in our numerical experiments making it difficult to set a stopping criterion. This requires the user to guess the number of iterations beforehand, which can greatly affect the quality of the final solution.

Additionally, a number of parameters must be set in both the PnP and MACE frameworks. PnP builds on either the ADMM or primal-dual algorithm, and hence inherits the user-defined parameters of such algorithms, which are either step-sizes or other hyperparameters that control the rate of convergence. Whilst for the primal-dual algorithm, there exist conditions for the step-sizes for the algorithm to converge, these conditions no longer guarantee convergence for PnP. \cite{Zhang2017b} have shown that the DRUNet network is best suited for PnP algorithms, since it has an additional channel that represents the noise level in the data, and is again user-defined. MACE also has a number of user-defined parameters, one related to the network, the regularization parameter of the inversion, and the weights of equation \eqref{eq:weighted_sum}. These parameters are not intuitive to set and we found that they can have a large influence on the final inversion results.

Finally, PnP can in principle also be used to perform 3D post-stack seismic inversion with 2D pre-trained denoisers, similar to MACE. Define
\begin{equation}
    \mathcal{J}(m) = \frac{1}{2}\min_m\Vert WDm - d\Vert_2^2 + g(Km),
\label{eq:prior_stacking_objective}
\end{equation}
with 
\begin{equation}
    g(Kx) = \sum_{i=1}^n g_i(K_ix)=\sum_{i=1}^n g_i(v_i), \quad K = \left[K_1;\:\ldots\:;K_n\right], \quad x = \left[x_1;\:\ldots\:;x_n\right]. 
\end{equation}
The proximal operator of $g$ is then given by \citep{Heide2014}
\[
    \text{prox}_g(v) = \left[\text{prox}_{g_1}(v_1);\:\ldots\:;\text{prox}_{g_n}(v_n)\right].
\]
Replacing all the proximal operators with denoisers allows one to apply PnP to 3D problems whilst still leveraging 2D pre-trained denoisers, in a similar spirit to MACE. We have tried implementing this strategy to compare it to MACE, however, the results were poor. Perhaps different parameter settings may lead to a result qualitatively more similar to MACE, however as this is a time-consuming process we feel that this result would not contribute that much to the current work. Lastly, in this work we have used MACE for the specific task of denoising a 3D volume using 2D denoisers, but it should be understood that MACE is a flexible, general-purpose method that allows the use of different agents to regularize the inverse problem at hand.

\section{Conclusions}
In this work, we have proposed the use of PnP for 2D seismic inversion and MACE for both 2D and 3D seismic inversion using 2D pre-trained deep denoisers. Our numerical examples on a 2D section of the 3D SEAM model reveal that the PnP methodology can achieve high-quality results, outperforming state-of-the-art Total-Variation regularization. More specifically, PnP is able to better resolve the complex layers of the subsurface than TV regularization, yielding cleaner and better resolved acoustic impedance models from seismic migrated data. Interestingly, the SNR for the TV solution is higher, indicating that SNR alone is not a good metric for seismic inversion. As far as 2D inversion is concerned, MACE is shown to produce results that are almost identical to those of PnP, despite a marginally higher SNR.

Additionally, we have shown that MACE can achieve state-of-the-art results for 3D post-stack seismic inversion using 2D pre-trained denoisers. A lack of availability of pre-trained 3D denoisers prohibits a direct extension of PnP to 3D, whilst MACE fills this gap. Although MACE should achieve consensus equilibrium, meaning that all agents produce the same solution at convergence, we found that this is not achieved in our experiments. Interestingly, the lack of consensus equilibrium does not seem to have a direct relation with the SNR of the solution, since the point where the consensus curves start diverging does not correspond to a sudden decrease in SNR. This result suggests that even though MACE does not actually achieve consensus equilibrium, it still produces high-quality solutions for the post-stack seismic inverse problem.
\bmhead{Acknowledgments}
The authors would like to thank KAUST and the sponsors of the Deepwave consortium for sponsoring this research.

\section*{Declarations}

\begin{itemize}
\item Funding: This work is funded by King Abdullah University of Science and Technology (KAUST) and the DeepWave Consortium.
\item Conflict of interest: not applicable
\item Availability of data and material: data are available upon request.
\item Code availability: code is currently not available.
\end{itemize}

\bibliography{main_article}

\end{document}